\documentclass[sn-aps,iicol]{sn-jnl}

\usepackage[T1]{fontenc}
\usepackage[utf8]{inputenc}
\usepackage{soulutf8}
\usepackage{soul}
\usepackage[dvipsnames]{xcolor}

\usepackage{hyperref}
\usepackage{caption}
\usepackage{subcaption}
\usepackage{geometry}
\usepackage[export]{adjustbox} 
\usepackage{multicol}
\usepackage{blindtext}
\usepackage[font=small,labelfont=bf]{caption}
\usepackage{siunitx}
\usepackage{booktabs, makecell, multirow}

\usepackage{hyperref}

\definecolor{myblue1}{rgb}{0, 0.2, 1}

\usepackage{graphicx}%
\usepackage{multirow}%
\usepackage{amsmath,amssymb,amsfonts}%
\usepackage{amsthm}%
\usepackage{mathrsfs}%
\usepackage[title]{appendix}%
\usepackage{xcolor}%
\usepackage{textcomp}%
\usepackage{manyfoot}%
\usepackage{booktabs}%
\usepackage{algorithm}%
\usepackage{algorithmicx}%
\usepackage{algpseudocode}%
\usepackage{listings}%

\theoremstyle{thmstyleone}%

\theoremstyle{thmstyletwo}%

\theoremstyle{thmstylethree}%

\begin{document}

\title{ Unlocking extreme doping and strain in epitaxial monocrystalline silicon }

\author*[1]{\fnm{Léonard} \sur{Desvignes}}\email{leonard.desvignes@hotmail.fr}
\author[1]{\fnm{Dominique} \sur{Débarre}}
\author[1]{\fnm{Ludovic} \sur{Largeau}}
\author[1]{\fnm{Géraldine} \sur{Hallais}}
\author[1]{\fnm{Gilles} \sur{Patriarche}}
\author[3]{\fnm{Giacomo} \sur{Priante}}
\author[3]{\fnm{Eric} \sur{Ngo}}
\author[1]{\fnm{Olivia} \sur{Mauguin}}

\author[2]{\fnm{Alberto} \sur{Debernardi}}
\equalcont{Corresponding author for first-principles calculations : {alberto.debernardi@mdm.imm.cnr.it}}
\author[3]{\fnm{Bernard} \sur{Sermage}}
\author*[1]{\fnm{Francesca} \sur{Chiodi}}\email{francesca.chiodi@universite-paris-saclay.fr}

\affil*[1]{\orgdiv{Centre de Nanosciences et de Nanotechnologies}, \orgname{Université Paris-Saclay, CNRS}, \orgaddress{\street{10 Boulevard Thomas Gobert}, \city{Palaiseau}, \postcode{91120}, \country{France}}}

\affil[2]{\orgdiv{Sede Agrate Brianza}, \orgname{CNR-IMM}, \orgaddress{\street{via Olivetti 2}, \city{Agrate Brianza}, \postcode{I-20864}, \country{Italy}}}

\affil[3]{\orgname{Probion Analysis}, \orgaddress{\street{Bat Alpha, 3 avenue du Canada}, \city{Les Ulis}, \postcode{91490},  \country{France}}}


\abstract{Hyperdoping, overcoming the solubility limit of dopants in a crystalline semiconductor, is a fertile method for the enhancement of the electrical, structural and optical devices' performances and for the exploration of exotic phases such as superconductivity. We demonstrate an unprecedented control on the dopant concentration and lattice deformation via nanosecond laser doping in epitaxial boron doped silicon, achieving record active concentrations ($8\,at.\%$) and lattice deformations ($3\,\%$). Probing the microscopical hyperdoping limitations, we show that the relevant mechanisms are caught by a simple combinatorial model, which quantitatively explains both the experimental carrier concentration and lattice deformation evolution. First principle calculations complete and support such simple model. Indeed, at the high doping levels now attainable, the maximum carrier concentration is inherently limited by the probability of two or three substitutional dopants occupying neighbouring lattice sites, forming partially inactive complexes that we detail. This description is valid in the case of perfect layers with no crystallographic defects and a fully substitutional dopant occupation, highlighting the quality of the epitaxial layers realised. }

\keywords{Semiconductors, hyperdoping, atomic-defect, electrical activation, lattice deformation}



\maketitle

\section*{Introduction}\label{sec1}
Hyperdoping, introducing impurity concentrations above the solubility limit in a semiconductor crystalline lattice, is a fertile method to tune widely the structural, electrical, and optical properties. Hyperdoping can give rise to new phases, such as superconductivity in Si and SiGe ultra-doped with B \cite{Bustarret2006,Nath2024}. It allows tuning the material properties into ranges of interest for applications, such as surface plasmons brought to the mid-infrared range and telecom wavelengths in ultra-doped Si:B and Si:P \cite{Poumirol2021}, for chemical/ biological sensing and thermal imaging \cite{Jiang2022}. Ultra-doping is also essential to answer the growing demands of digital technologies in terms of information processing capacity and frequency. Indeed, while better performances are attained by reducing the transistor dimensions \cite{IRDS2022}, this in turn leads to the increase of the metal-semiconductor source/drain contact resistance, the dominant parasitic factor \cite{raghavan2015}. Hyperdoped semiconductors allow instead achieving record contact resistances $\mathrm{R_{int}.A \sim 10^{-7} - 10^{-9}\,\Omega.cm^2}$ \cite{chiodi2014, chiodi2017, gallacher2012, everaert2017, hung2018, wang2017}, if the dopant concentrations are high enough (in Si, a dopant concentration of 2 at.$\%$ corresponds to  1 charge/$\mathrm{nm^3}$).

The combination of ion implantation with thermal annealing, as well as conventional epitaxy growth techniques, can attain high doping levels exceeding the solubility limit ($n_{sol} \sim 1\, at.\%$), but only with partial activation of the dopants that rather form aggregates or precipitates
In order to obtain higher charge concentrations and activation efficiencies in higher quality layers, non-equilibrium processes are the key. Pulsed laser melting epitaxy (PLME) is a well-established method that takes advantage of ultra-fast growth regimes. A short laser pulse ($\mathrm{t\sim ns}$) is used to locally melt the substrate, which then recrystallizes epitaxially at a few ${m/s}$. In this fast liquid phase epitaxy, the dopants do not have the time to diffuse to their equilibrium configuration and a meta-stable ultra-doped layer is formed, with full activation well above the solid solubility limit. PLME has extensively been applied to obtain high doping levels and sharp, box-like profiles in group IV semiconductors doped with III \cite{young1978a,kerrien2002,milazzo2020,wang2017} and V \cite{young1978b,williams1982,huang2005} elements, as well as chalcogens \cite{li2017,winkler2011,winkler2012,gandhi2020a,zhou2015a,wang2018,wang2019a,wang2019b,wang2020} and transition metals \cite{yang2019,gandhi2020b,dissanayake2022,lim2021,dong2023,wen2021}.

The question is now of how far PLME can be pushed in terms of extreme active concentrations. Several theoretical and experimental studies have shown that the carrier density of doped semiconductors are inherently limited by the formation of atomic defects, such as vacancy-impurity complexes \cite{pandey1988, mueller2003} or dimers \cite{mueller2004,voyles2002,voyles2003} introducing localized deep-level states in the band gap of n-type Si \cite{chadi1997}. Few-atoms defects play a role in Si doped with transition metals \cite{yang2019} and chalcogens \cite{debernardi2021,wang2019a}. Yet, the limitation of the electrical activation in p-type Si has systematically been attributed to the formation of precipitates,
raising the question of the possibility of higher levels of activation in p-Si by optimizing the doping process.\\

In the current work, we achieve state of the art hole concentrations, dopant activation levels, and high tensile strain in monocrystalline boron doped Si epilayers, synthesized by Gas Immersion Laser Doping (GILD) under ultra-high-vacuum.
Hall effect and Secondary Ion Mass Spectrometry (SIMS) are used to study the evolution of the carrier density $h$ with respect to the total B concentration $C_B$. Furthermore, we analyse the structural defects and the evolution of the lattice deformation due to the B substitutional epitaxial incorporation by X-Ray Diffraction (XRD) and Scanning Transmission Electron Microscopy (STEM).\\
We demonstrate hole concentrations as high as $h=8\,at.\%$ ($h=4\times 10^{21}\,cm^{-3}$), using a Hall coefficient $\gamma = 0.7$, with lattice deformations up to $3\,\%$ in monocrystalline layers 20 to 315 nm thick. To investigate the microscopic origin of the saturation of both charge carrier density and lattice deformation, we quantitatively explain our data with a simple binomial model. Indeed, at high dopant concentration, the probability that two or more B atoms occupy neighbouring lattice sites, leading to partially inactive complexes, cannot be neglected, pointing to an intrinsic 'geometrical' limit to the maximum hole concentration achievable.
With a more in-depth analysis, we confirm the conclusions of this immediate model with the results of first principle DFT simulations calculating in a random dopant distribution the formation energy of multiple types of defects formed by up to 3 boron atoms. We find an excellent agreement if we allow for a small dopant mobility around the lattice site, namely between the first and second nearest neighbours. The same model allows to explain quantitatively the lattice deformation, which provides independent information on the inactive defects, not participating in the carrier density but contributing to the lattice deformation.
In conclusion, this work provides accurate predictions of SiB activation and structure at previously unattained extremely high concentrations, demonstrating a very general inherent activation limitation in group IV semiconductors that can only be observed upon very fast anneal, with conclusions that may be extended to other semiconductors and dopant atoms. 

\section*{Synthesis of hyperdoped Si:B epilayers}\label{sec2}
\begin{figure}[t]
    \centering
    \includegraphics[width=1\linewidth]{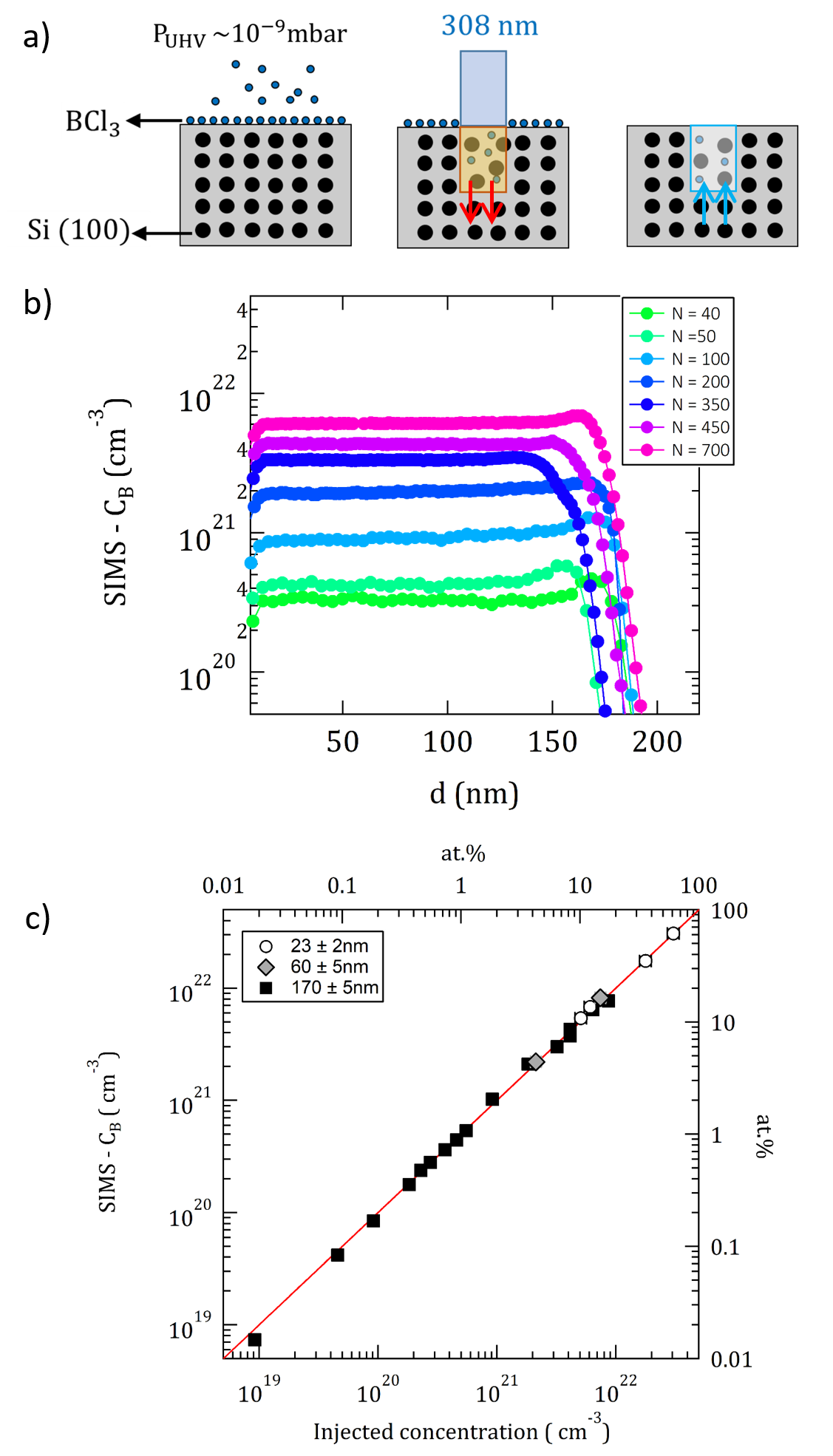}%
    \caption{Doping of hyperdoped SiB by GILD: (a) Gas Immersion Laser Doping : (i) chemisorption of (BCl\textsubscript{3}) gas on the Si surface, (ii) pulsed laser melting inducing a downwards melting front (iii) epitaxial rapid solidification trapping B atoms within the crystalline lattice. (b) SIMS concentration profile of a layer $170\pm5nm$ thick for various number of laser shots N. (c) Atomic concentration $C_B = \frac{D_B}{d}$, with $D_B$ the integrated dose of the SIMS profiles and d the thickness, vs the injected concentration at each laser shot $C_{inj}\sim \frac{D_{surf}\times N}{d}$. The red line highlights the perfect equality between $C_B$ and $C_{inj}$.}
    \label{fig:fig1}
\end{figure}

Gas Immersion Laser Doping is performed in a reactor with ultra high vacuum (residual level $\sim 10^{-9}$ mbar). A puff of boron trichloride precursor gas (BCl\textsubscript{3}) is injected over the surface of a (100) oriented n-type silicon ($\rho=80\; \Omega.cm$) with sufficiently high pressure to saturate the chemisorption sites (see Fig.\ref{fig:fig1}a). The B atoms are then incorporated into the system by melting the silicon with an excimer laser pulse ($\lambda = 308\,nm$, $25\,ns$ pulse)  with fluency ranging from $E= 0.9$ to $1.8 \;$J/cm$^2$, allowing the control of the melted depth $d$ from 20 to 315 nm, with a few nm precision, as measured by STEM \cite{hallais2023} and SIMS (see Methods for the determination of the thickness). 
In the liquid phase, boron diffusion is very fast ($D \sim 10^{-4} cm^2/s$) as compared to its diffusion in solid Si near the melting temperature ($D \sim 10^{-11} cm^{2}/s$) \cite{kodera1963}, so that the boron concentration is limited to the melted region, where it is homogeneously distributed. At the end of the laser pulse, the fast cooling induces a fast epitaxy ($\sim 4 m/s$ \cite{wood1981}), allowing for a segregation coefficient $k$ close to 1 \cite{wood1981}, so that the boron present in the liquid phase is in practice totally incorporated in the solid phase. This induces a homogeneous boron profile in the epilayer, with a sharp Si:B/Si interface. \\
The total atomic concentration profile $C_B (d)$ is measured by SIMS, with particular care taken to account for matrix effects at such high doping levels (see Methods). In Fig\ref{fig:fig1}b and Fig.\ref{fig:fig3}b, SIMS concentration profiles and TEM images show the excellent vertical homogeneity of the doping, with $ \frac{\delta C_B}{C_B} =13\pm8\;\%$ and a few nanometers wide interface. A slight accumulation of B atoms is observed at the SiB/Si interface. In the most doped samples, this accumulation is composed by inactive B, as confirmed by TEM images showing in the same thickness range the presence of B aggregates.  Such B accumulation accounts at most for $1\%$ of the total B concentration up to $C_{B}=7\times10^{21}cm^{-3}$ ($N=700$ laser shots).  \\
Each laser doped area is subjected to several laser pulses $N= 1-700$ that allow a fine tuning of the doping level as each shot introduces a fixed surface density determined by the density of the saturated chemisorption sites. This is shown in Fig.\ref{fig:fig1}b where the SIMS concentration increases proportionally with the number of laser shots. The total dopant concentration $C_B$ is given precisely by the amount of total injected dopants, up to surprisingly large concentrations ($\sim 60 at.\%$), with no observable desorption due to incorporation saturation,
as shown in Fig.\ref{fig:fig1}c.  \\

\section*{State of the art hole concentration}\label{sec4}

\begin{figure}[t]
    \centering
    \includegraphics[width=\linewidth]{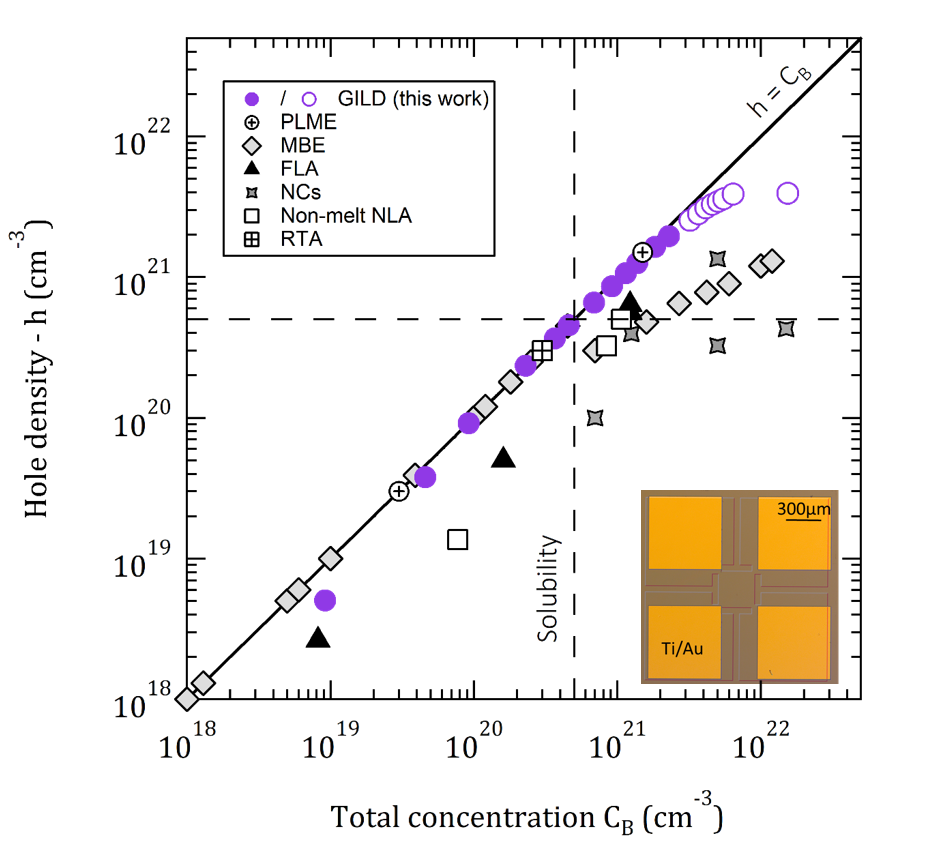}%
    \caption{Comparison of the carrier density vs the B atomic concentration achieved by various annealing methods : white crossed squares : RTA \cite{solmi1990}; black triangles : FLA \cite{jain2004,yeong2008,nishikawa2011,do2014}; white squares : non-melt NLA \cite{earles2004,cristiano2016,sharp2006}; dark grey stars : NCs \cite{zhou2015b,patra2019,hunter2019} ; grey diamond : MBE \cite{glass2000} ; purple circle : GILD (this work). For the latter two doping techniques, filled markers are attributed to pseudomorphic grown layers and unfilled markers to (partially) relaxed layers. The inset image shows a picture of the Hall cross geometry.}
    \label{fig:fig2}
\end{figure}

The hole concentration $h$ is probed by Hall measurements on a 300$\,\mu$m square Hall cross etched in the laser-doped spot and contacted with Ti/Au electrodes (see Methods). 
Fig.\ref{fig:fig2} reports the highest hole concentrations $h$ vs atomic concentrations $C_B$ achieved in heavily doped Si:B layers by various annealing and doping techniques. Except for nanocrystals (NCs), the active dose is systematically evaluated by Hall measurements in these works. It can be seen that the evolution of $h(C_B)$ and the activation efficiency differ from one doping technique to another, suggesting specific deactivation processes for each one of them.

Rapid thermal annealing (RTA), flash lamp annealing (FLA) and non-melt ns laser annealing (non-melt NLA) techniques involve a pre-implantation step to introduce the dopants, which is responsible for the formation of deep Si interstitials (I-defects). During the annealing process, these defects enhance the diffusion of B atoms and trap them to form electrically inactive boron interstitial clusters (BIC) \cite{jain2002}. A high thermal budget annealing, such as RTA ($t \sim s$) and conventional furnace anneal ($t \sim \mathrm{hours}$), is required to completely get rid of these impurities \cite{aboy2014} but causing an out-diffusion that dilutes the system below the solid solubility. FLA ($t \sim ms$) and non-melt NLA ($t \sim ns$), on the other hand, suffer from their low thermal budget, responsible for the formation of larger and more stable I-defects that prevent a full activation \cite{rebohle2019,marques2021}, with e.g. 50$\%$ activation just above the solubility limit for $h=6.4\times 10^{20}$cm$^{-3}$    (Fig.\ref{fig:fig2}).

The deactivation process is different when using growth techniques. For example, gas-source molecular-beam epitaxy (GS-MBE) benefits from the passivation of the Si surface from hydrogen atoms, present in the $\mathrm{Si_2H_6}$ and $\mathrm{B_2H_6}$ molecular beams, which favors the incorporation of B atoms within the bulk rather than the spontaneous accumulation on the surface \cite{luo2003}. However, the activation remains partial above $C_{sol}$, with 24$\%$ activation for $h=6.4\times 10^{20}$cm$^{-3}$ and down to $11\;\%$ for the maximum hole concentration achieved $h\sim 1.3\times10^{21}$cm$^{-3}$, because of the formation of neutral B pairs incorporated as a unit in substitutional sites. Such defects are also suspected in reduced pressure - chemical vapor deposition (RP-CVD) samples grown with the same precursor gases \cite{hartmann2020}. On the other hand, nanocrystals are mostly believed to be electrically limited by the accumulation of B atoms on the surface \cite{oliva2016}.\\

\begin{figure*}[t]
    \centering
    \includegraphics[width = 1\textwidth]{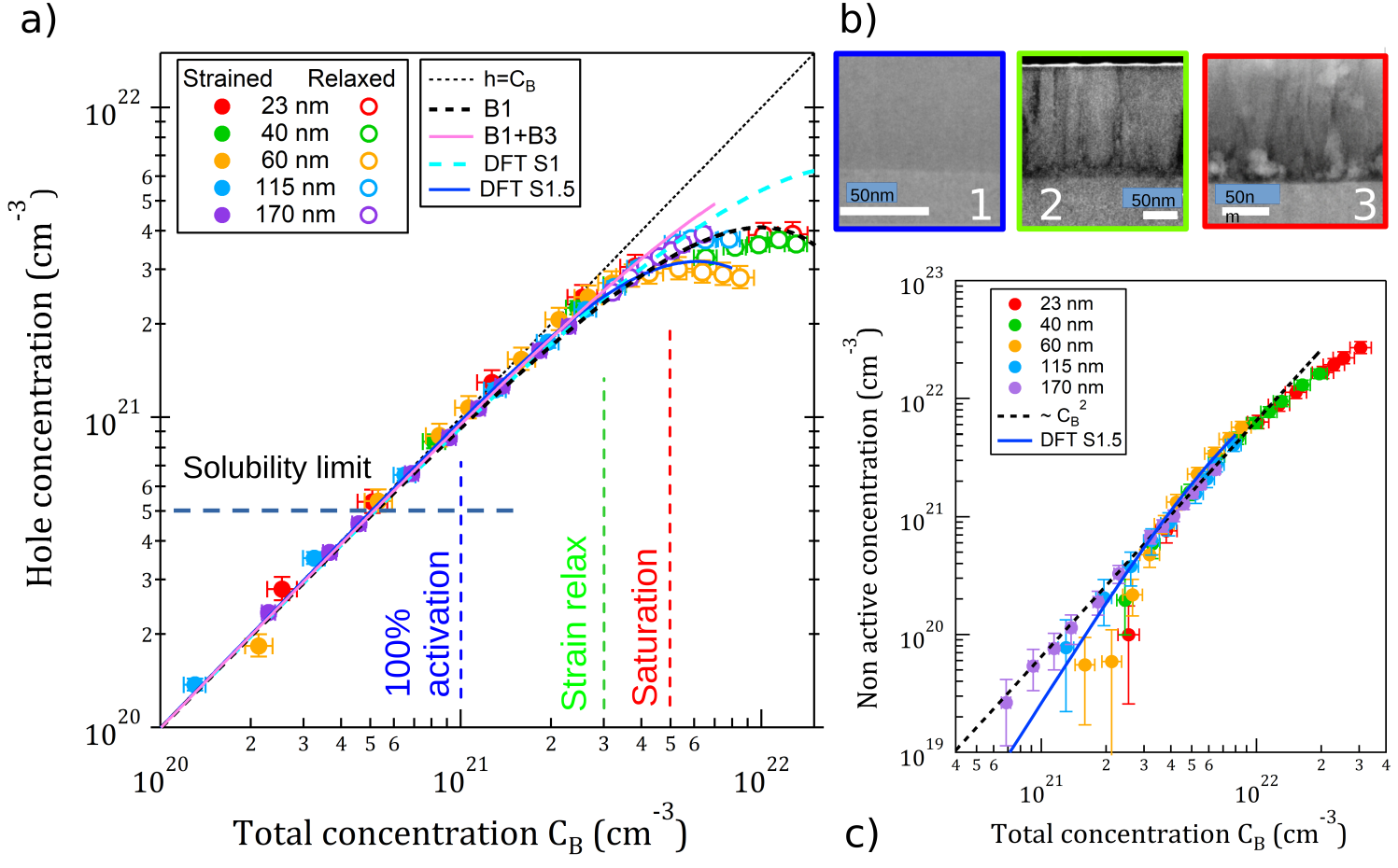}
    \caption{(a) $\mathrm{h(C_B)}$ of hyper-doped SiB layers synthesized by GILD with thickness $d=23$ to $170\,$nm.  The vertical dashed lines highlight the 100$\%$ activation concentration range, the onset of strain relaxation and the onset of saturation. As the relaxation onset slightly depends on the thickness, and for clarity, for each series pseudomorphic layers are shown in full circles, relaxed ones in empty circles. The fine black line represents the evolution expected for 100$\%$ activation ($h=C_B$).
    Four theoretical models are compared to the experiments: the concentrations of B monomers (black dotted line B1, eq.\ref{eqB1}) and B monomers and trimers (pink dotted line, B1+B3) expected from a simple binomial model, and the first principle simulations accounting for complexes with 1 to 3 B atoms for the diffusion shell S1 (DFT S1, light blue dotted line) and S1.5 (DFT S1.5, blue line), discussed in sec. \ref{dft}.  (b) Bright field STEM images on top show: (left) a fully strained pseudomorphic layer, perfectly crystalline with no defects; (middle) a partially relaxed layer with B-poor pile-up regions; (right) a fully saturated layer showing evidence of B aggregates at the interfaces. The B-poor pile-up regions are due to cellular breakdown, where relaxation process induce the appearance of compositional column-shaped defects at high crystallisation rates and high concentrations. (c) Non-active concentration $ C_{NA} = C_B - h$ vs total concentration $C_B$ compared to a quadratic power law (dotted black line) and to the expectations of the first principle simulations for complexes with 1 to 3 B atoms and the diffusion shell S1.5 (blue line).    }
    \label{fig:fig3}
\end{figure*}

The evolution of $h(C_B)$ in samples synthesized by GILD or PLME stands out from the other methods, suggesting an alternative deactivation mechanism. In our study, $h$ increases continuously with an activation rate of 100$\%$ up to $1\times10^{21}cm^{-3}$,  more than twice the solubility limit, and still as high as 60$\%$ at the maximum hole concentration attained $h_{sat} = 8 \,at.\% = 4\times10^{21}cm^{-3}$, with Hall coefficient $\gamma = 0.7$, the state of the art for the carrier concentration in silicon (Fig.\ref{fig:fig2}).
Overall, this evolution does not depend on the thickness of the doped layer, except for a slightly higher saturation concentration for the thicker layers (see Fig.\ref{fig:fig3}a). \\
The saturation level marks an abrupt stop in the evolution of $h(C_B)$, which is a sign that the system has reached its \emph{non-equilibrium} solubility limit. This has been actually observed in STEM analysis with the formation of B-rich precipitates localized at the interfaces of the layer \cite{hallais2023} (see also right STEM image in Fig.\ref{fig:fig3}a)). 
The transition from a pseudomorphic state to a partially relaxed state does not seem to affect the evolution of $h(C_B)$, which is continuous through the relaxation.

\section*{Ultra-doping: inherent activation limitation}\label{sec5}
To understand the limitation of the hole concentration and of the activation rate, it is necessary to recall the order of magnitude of the atomic B concentration : $C_B$ is increased up to 13 at.$\%$ before attaining the saturation. In this regime, B atoms are randomly distributed, with the absence of any cluster of more than 6 B atoms, as observed by Atom Probe Tomography \cite{hoummada2023}. Because of the high B concentrations, there is a non-negligible probability that two or more B atoms find themselves located in neighboring lattice sites during the crystallization process, making it possible to form inactive complexes. 
In a very intuitive model, we have calculated the probability to find single, isolated, active, substitutional B atoms surrounded by 4 Si atoms. This probability is given by the binomial distribution 
\begin{equation}
 B_1 = C_B \times  \left ( 1 - p\right)^4
\label{eqB1}
\end{equation}
where $p= C_B/ n_{SiB}$ and $(1-p)$ are respectively the probability of picking one B or one Si atom among all the available atoms and $B_1$ the B monomers concentration. For simplicity, $n_{SiB}$ is fixed as $n_{Si}=5\times10^{22}cm^{-3}$, the intrinsic atomic density of c-Si, as the variations induced by the lattice deformation under strain are evaluated at most a few percent. 
The resulting hole concentration associated to the B monomers is shown in Fig.\ref{fig:fig3}.a. Strikingly, it reproduces quantitatively, extremely well, the experiment, keeping in mind the simplicity of the model : for the $d=170\,$nm series, we observe less than $3\%$ relative difference with the experiments up to $\sim 3\times10^{21}cm^{-3}$, and at most $5\%$ over the whole range. As the saturation regime is approached, aggregates  start appearing (Fig.\ref{fig:fig3}b) and this picture is no longer valid.
The conclusions of this first simple model are very strong. An intrinsic limit to the maximum hole concentration achievable is predicted, estimated here at $\sim 4.1\times10^{21}cm^{-3}$, even with substitutional, $100\%$ active dopants, in the complete absence of defects. This limit is simply due to the fact that at extremely high concentrations, close to the concentrations now possible experimentally, the probability of finding two or more B atoms in neighbouring sites is not negligible anymore.\\

The second strong conclusion is that what limits the further increase of the hole concentration is the formation of complexes formed by a few B atoms. Fig. \ref{fig:fig3}.c plots the evolution with the total concentration $C_B$ of the non-active B atoms concentration, $C_{NA} = C_B -  h$. Quite strikingly, $C_{NA}$ increases nearly perfectly quadratically with $C_B$ up to the saturation concentration ($C_B \sim 5-7 \times 10^{21}cm^{-3}$), pointing towards 2-atoms B complexes, B dimers, as the main responsible of the activation limit. Around the saturation concentration, a kink is visible highlighting a change in the power law describing the evolution of the inactive concentration, as larger complexes, up to a few tens of nanometers aggregates, appear.

\subsection*{What are the inactive complexes?}\label{subsec4_1}

\begin{figure}[t!]
    \centering
    \includegraphics[width=2.5\linewidth, angle=270]{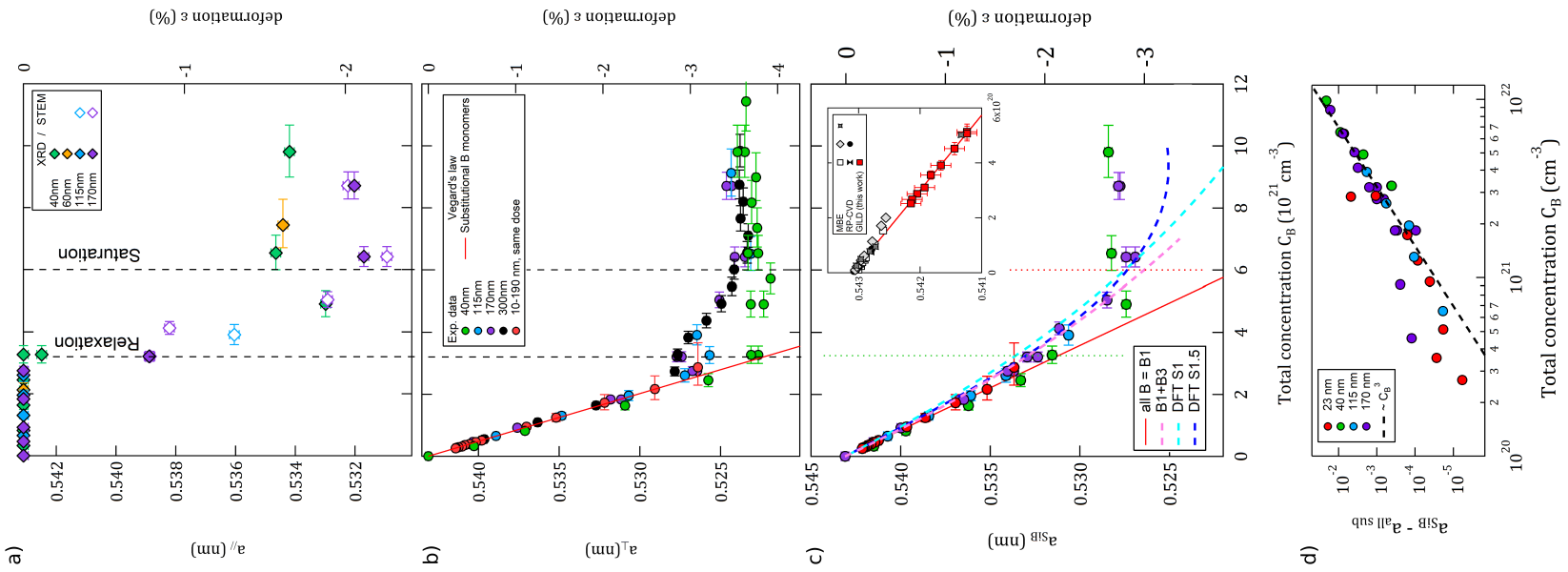}
    \caption{(a) In-plane lattice parameter $a_{//}$, extracted from both XRD-RSM maps (filled markers) and STEM-GPA images (unfilled markers). (b) Out-of-plane lattice parameter $a_{\perp}$ (XRD-RSM maps only). The red line has a slope of $\sim - 6.5\times10^{-24}\,$nm$\,$cm$^3$. (c) Lattice parameter $a_{SiB}$. The inset shows a closer view over the first $6\times10^{20}cm^{-3}$ and compares this work to MBE \cite{vailionis1999, sardela1994, baribeau1991} and RP-CVD \cite{boureau2018,hartmann2020} growth techniques. Each data item corresponds to fully activated layers. (d) Difference between the measured lattice parameter and the one expected for a fully monomer concentration, compared to a third power law.}
    \label{fig:fig4}
\end{figure}
While the evolution of the non-active concentration suggests the formation of 2-atoms complexes as main deactivation mechanism, it is difficult to identify which one(s) of the possible configurations involving two atoms dominate.
However, while inactive, different complexes contribute differently to the lattice deformation of the SiB layer. In order to better understand and lift the degeneracy of the different contributions, we have investigated the structural properties by means of XRD-reciprocal space maps (XRD-RSM) measurements around the (004) and (224) reciprocal points as well as geometrical phase analysis (GPA) on STEM images.   
The lattice parameters in-plane ($a_{//}$) and out-of-plane ($a_{\perp}$) are extracted from the maps and their evolution with $C_B$ are respectively shown in Fig.\ref{fig:fig4}a and Fig.\ref{fig:fig4}b for samples identical to those measured in Fig.\ref{fig:fig3}, with two additional datasets: a series with $d=300\,$nm and a series with a fixed doping surface dose ($4.6\times10^{15}cm^{-2}$) and varying thickness. For each dataset, the first $2\times10^{21}cm^{-3}$ are characterized by a linear decrease of $a_{\perp}$ while $a_{//}$ remains equal to $a_{Si}=0.5431\,$nm, characteristic of dislocation-free pseudomorphic layers with a maximum (state of the art) deformation of $\delta a_{\perp}/a_{\perp} =-3.7\%$. At higher concentrations, $a_{//}$ decreases with $C_B$ and then saturates at $C_{sat}$, while $a_{\perp}(C_B)$ either saturates for the thinnest layers ($40nm$) or has a local maximum followed by saturation. These events mark the beginning of plastic relaxation and are consistent with the results of Ref.\cite{hallais2023}. Note that $a_{sat_{//}} > a_{sat_{\perp}}$: the relaxation is only partial.  The slight increase of $a_{\perp}$ at the highest doping is in line with Ref.\cite{bisognin2006a}, reporting an expansive effect of the crystal lattice with the appearance of B clusters in the saturation regime.

We calculate from $a_{//}$ and $a_{\perp}$ the relaxed lattice parameter of SiB, $a_{SiB}$, supposing an isotropic material: 
\begin{equation}\label{aSiB_isotropic}
    \frac{a_{\perp}-a_{SiB}}{a_{SiB}} = K\cdot\frac{a_{//}-a_{SiB}}{a_{SiB}}    
\end{equation}
where $K=\frac{-2\nu}{1-\nu} \simeq -0.77$. The Poisson coefficient is fixed to $\nu = 0.278$, keeping the same value for SiB than for Si, as the first principle calculations performed predict less than a few percent variations between them. According to Vegard's law, $a_{SiB}$ is expected to evolve linearly with the substitutional fraction $n_s$ of B atoms:\\
$a_{SiB} = x_s\cdot a_B + (1-x_s)\cdot a_{Si}$\\
where $a_B$ is the lattice parameter of an ideal diamond-cubic B crystal. This is indeed the case over a concentration range where most of the dopants are in substitution ($\sim 95 \%$ at $C_B = 6\times 10^{20}cm^{-3}$), as shown in the inset of Fig.\ref{fig:fig4}c, where GILD, CVD and MBE layers are compared.
 In this range, all the experimental data show a common linearity given by $a_B = 0.36\pm0.015\,$nm. This value is slightly lower than the experimental $0.378\pm0.007\,$nm found in Ref.\cite{bisognin2007} but is in good agreement with the ab.initio value $a_B \simeq 0.358nm$ \cite{dunham2006}. Our first principles simulations predict $a_B = 0.375$ nm.  
Over a wider concentration range, $a_{SiB}$ gradually deviates with $C_B$ and saturates ($a_{sat}$) at $C_{sat}$. Only considering the contribution to the lattice deformation of the B monomers cannot account for the experimental results anymore, as the strain associated to B complexes slows down the deformation evolution with $C_B$ (Fig.\ref{fig:fig4}).
However, the experiments can be still described with Vegard's law  at high concentrations if the lattice deformation induced by the complexes is taken into account:
\begin{equation}\label{eq_Veg_extd}
    a_{SiB} = \displaystyle\sum_{i} x_{i}\cdot a_{{i}}+(1-\displaystyle\sum_{i}x_{i})\cdot a_{Si}
\end{equation}
where $x_i$ is the concentration fraction of each of the $i$ complexes and $a_i$ their fictional lattice parameter. 
The lattice parameters of the main possible complexes were obtained through first principle calculations.

\subsection*{First principle simulations of the carrier concentration and lattice deformation}\label{dft}
To confirm and complete the interpretation of the binomial model, we carried out first-principles simulations to compute both the carrier concentration and the lattice deformation induced by charge-neutral B complexes formed by 1 to 3 B atoms. The probability of each complex was evaluated from its formation energy, calculated at high doping, with Boltzmann weights. Then,  each complex electrical activity and induced deformation was computed to model SiB hole concentration and lattice parameter.\\
\begin{figure}[t]
    \centering
    \includegraphics[width=\linewidth]{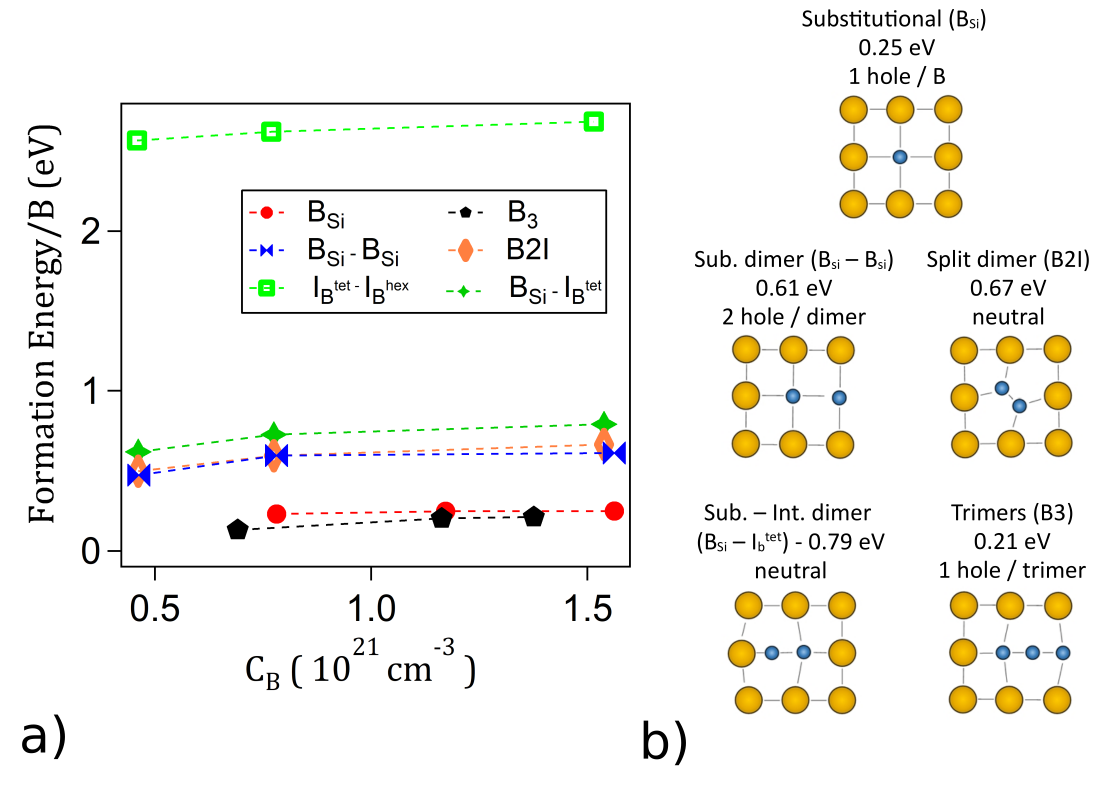}
    \caption{    (a)  First-principles simulations of the formation energy of distinct types of charge-neutral B complexes involving 1 to 3 atoms as a function of B doping concentration $C_B$. The probability of each complex is evaluated as a function of its formation energy with Boltzmann weights. Through the calculated complex electrical activity, the expected hole concentration is calculated.
    (b) Schematic representation - inspired from Ref.\cite{luo2003} - of the different B-complexes with the electrical contribution as well as their average formation energy.
    }
    \label{dft}
\end{figure}
We consider the following, more probable, B complexes (see Fig. \ref{dft}.b for a stick-and-ball model of their atomic configuration ): monomers of substitutional B (B$_{Si}$); dimers of substitutional B in neighboring sites (B$_{Si}-$B$_{Si}$); split dimers formed by two B symmetrically displaced at a substitutional site (B2I); two interstitial B, one in hexagonal and the other in tetrahedral position ($ I_{B}^{tet} - I_{B}^{hex}$); one substitutional B and one interstitial B in a tetragonal position ($ B_{Si} - I_{B}^{tet}$); and B trimers formed by two substitutional B atoms at adjacent lattice sites having an interstitial boron in between ($B3$). While complexes with a larger number of B numbers are bound to appear at higher concentrations, we show that nanosecond laser doped layers up to the saturation regime can be fully described by such a small number of B atoms complexes.  \\
 In Fig. \ref{dft}a, we display our first principles results for the formation energy as a function of the B concentration $C_B$. In the hyper-doped range of concentration considered ($1-3 at.\%$), the formation energies are a nearly flat function of $C_B$, suggesting that this trend can be extrapolated at higher concentration; a similar behavior has also been found in chalcogen hyperdoped Si \cite{wang2019b, debernardi2021, debernardi2022}. The more energetically favorable complexes  are the B monomer and trimer,  presenting a formation energy at least few hundreds of millivolts lower than the rest of the complexes.    Among the complexes involving two B, B dimers, $B_{Si}-B_{Si}$ (shallow double acceptors), and B2I split (electrically inactive) present the lowest, very similar, formation energies, while B interstitial-substitutional pairs, $ B_{Si} - I_{B}^{tet}$ (electrically inactive), have a formation energy $\sim$0.12 eV higher. In comparison, the complex formed by two interstitial B, $ I_{B}^{tet} - I_{B}^{hex}$ has an activation energy $\sim$1.9 eV higher than the other complex studied, and thus is energetically unfavourable and unlikely to form at these concentrations. B trimers (single acceptor) have the lowest formation energy among all the complexes investigated, even B monomers 
 , and are favoured if three B have the possibility to be located close enough. At relatively low concentrations ($C_{B}\sim 1\;at.\%$), if we assume a random distribution of dopants and null or small mobility, the average distance between B atoms is too large to promote the formation of trimers, and the dopant population is almost exclusively composed by B monomers. At higher concentrations, roughly above $\sim 3\times10^{21}\,$cm$^{-3}$, $B3$ becomes dominant.  \\
The relative probabilities among different types of complexes composed by the same number of B can be assigned according to a Boltzmann weight $\propto e^{- \Delta E^{Form.}_{D}/k_BT}$ where $k_B$ is the Boltzmann constant and $T = 1683K$ the melting temperature of Si \cite{debernardi2021}. For complexes involving two B atoms $B_{Si}-B_{Si}$ represents $\sim 42.5\;\%$ of the doublets, ${B2I}\sim 42.5\;\%$ and ${B_{Si}-I_B}\sim 15\;\%$. In the following models, the contributions of ${B_{Si}-I_B}$ and $ I_{B}^{tet} - I_{B}^{hex}$ are omitted because of the low populations
. Thus, on average, each B dimer provides approximately one hole to the valence band. When three B atoms are involved, only $0.05\;\%$ form $\mathrm{B_{Si}-B_{Si}}$ complex, while $90\%$ form of $B3$ complexes.\\
To further optimize this model, it is possible to assume a small diffusion of the dopants during or right after the crystallization process \cite{debernardi2021,debernardi2022}.  We consider the nearest neighbours shell of lattice sites (S1), the second nearest neighbours shell (S2), and the intermediate case S1.5, with half of the sites of the next nearest neighbor shell. The dopants are allowed to diffuse within the shell considered where they arrange according to the Boltzmann weigh computed by the formation energy of the complex.  For an intuitive picture, we can consider the first B to interact (by diffusion) up to another B located in the whole (S2) or partial (S1.5) next nearest neighbours shell, while for the S1 case the dopants are basically “frozen” in the lattice sites, allowing only the B doublets in nearby sites to arrange according to their relative Boltzmann weights.  
After the maximum hole concentration is reached, the possibility of clusters having size larger than three B is non-negligible; to account for this,  we evaluate the probability to have $m$ B, with $m>3$, within the shell considered, and  we assume that if more than three B are present they aggregate in an inactive cluster.  Note that this approximation is expected to influences significantly only the saturation regime, where we observe the formation of B clusters.\\
Excellent reproduction of the experimental data is achieved up to $\sim 2 ×\times10^{21} cm^{-3}$, for all models (S1, S1.5, S2). The best overall agreement is obtained, with no adjustable parameters, with S1.5 model, which accounts for a dopant diffusion comparable to the distance between two nearest neighbour lattice sites (Fig.\ref{fig:fig3}a). This low diffusion is compatible with the very short, sub-$\mu$s diffusion time allowed after the recrystallisation in the still-hot solid crystal: for 700 laser shots, and a cooling time of 100$\,$ns corresponding to our thicker layers, the solid-phase diffusion coefficient at the melt temperature ($D=4\times 10^{-11}\,cm^{2}/s$) gives an upper limit for the diffusion length of 0.5 nm.     \\

To model the lattice deformation, we consider the same complexes concentrations as for the $h(C_B)$ plot, and calculate the fictional lattice parameter associated to the effective strain induced by each complex.\\
Dimers of substitutional B are described by $a_{B_{Si}- B_{Si}}=0.380\,$nm, that we can safely assume equal within the numerical error to $a_{B_{Si}}=0.375\,$nm; for split dimers $a_{B2I} $= 0.552 nm, while for trimers $a_{B3}=0.490$ nm. Experimental studies conducted on the effect of cluster and $\mathrm{B2I}$ defects found that $\mathrm{B2I}$ induce a volume contraction $26\;\%$ smaller than $\mathrm{B_{Si}}$ ($a_{B2I}\sim 0.46nm$) \cite{bisognin2006a,bisognin2006b}. The comparison of our simulations with the experimental data in Ref.\cite{bisognin2006a, bisognin2006b} suggests that the B clusters individuated might be associated to B trimers rather than B2I dimers. In this case, both $a_{B_{Si}}$ and $a_{B3}$ numerical values are coherent with a $\sim 4\%$ overestimation of the experimental values. Such overestimation is a well-known effect associated to GGA (Generalized Gradient Approximation) underbinding tendency [26].
In Fig.\ref{fig:fig4}c the experimental lattice parameter is compared to the lattice deformation expected from our model. We recall that it describes a dimer population made up of $50\%$ substitutional dimers $\mathrm{B_{Si}}-\mathrm{B_{Si}}$ and $50\%$ split dimers $\mathrm{B2I}$. The contribution of $\mathrm{B_{Si}}-\mathrm{I_{B}}$ and $\mathrm{I_B}$ defects are omitted because of their low proportions.  
Both the fully strained and relaxed region are well reproduced. The compressive contribution of the increasing aggregates is also qualitatively correctly predicted, even if the model tends to underestimate the experimental data at the higher end of the saturation range. As the ab-initio simulations are applied to a fully isotropic SiB system, these discrepancies might be associated to the only partial relaxation of the Si:B layers.  
Like for the $h(C_B)$ curve, the best agreement is obtained allowing a small B diffusion, namely within the S1.5 shell, confirming the hypothesis of slight diffusion of the dopants.\\

The interpretation given above,  within an equilibrium model, of $h(C_B)$ and $a_{SiB}(C_B)$ experiments suggests that with Gas Immersion Laser Doping only B monomers ($B1$), $100\%$ active, are incorporated up to $\sim 1.5 ×\times10^{21} cm^{-3}$, well above the solubility limit. Then, progressively, dimers are formed, their population being composed of approximatively half substitutional, fully active, dimers ($B_{Si}-B_{Si}$), and half, inactive, split dimers ($B2I$), which contribute decreasing the overall tensile strain. Finally, trimers ($B3$) appear and dominate the evolution of the lattice deformation. This is highlighted by Fig.\ref{fig:fig4}d where the correction to a fully substitutional, Vegard's law, dependence $a_{SiB} (C_B)$ ($\Delta a = a_{SiB} - [(n_B/n_{Si})\, a_B + (1-n_B/n_{Si})\,a_{Si}]$) follows a third power law of the dopant concentration.\\
While this picture is very satisfying, one might wonder on the possibility of a different, out of equilibrium, microscopic interpretation, associated to the extreme recrystallisation speed of nanosecond laser doping. Our analysis shows that only one other description might explain both $h(C_B)$ and $a_{SiB}(C_B)$ at the same time. In this second description, monomers also dominate up to $\sim 1.5 ×\times10^{21} cm^{-3}$, but rapidly the triplets take over, as the more favorable complexes at high concentration. Fig.\ref{fig:fig3}a and Fig.\ref{fig:fig4}c display this simple description of a binomial monomer concentration (eq.\ref{eqB1}), with the rest of the dopants forming triplets ($B1+B3$), valid before the saturation threshold. This picture describes slightly better the evolution of the lattice deformation, especially for the thinnest layers, but slightly overestimates the carrier concentration approaching saturation. This highlights the central role of trimers, and suggests that the recrystallisation velocity might play a role in the microscopic dopant complexes formation, so that a faster nanosecond anneal (not too rapid to still allow a proper recrystallisation), might still allow a $\sim 10-15\,\%$ carrier concentration increase, ultimately 'geometrically' limited by the formation of larger complexes.



\subsection*{Conclusions}
Controlling the formation of small atomic complexes in p-type semiconductors and understanding their role in the carrier density limitation and strain relaxation is of the uppermost importance. Nanosecond laser doping, with its unrivalled growth rates, allows exploring this high carrier concentrations regimes. We discuss the case of boron-doped silicon, demonstrating state of the art hole concentrations as high as $h=8\,at.\%$ ($h=4\times 10^{21}\,cm^{-3}$), using a Hall coefficient $\gamma = 0.7$, and lattice deformations up to $3\,\%$ in monocrystalline epitaxial layers 20 to 315 nm thick.
Thanks to a model based on a simple binomial law, assuming all dopants substitutional, we are able to quantitatively predict the evolution of the carrier density and the lattice parameter at unprecedented concentrations. We demonstrate that the intrinsic limited is associated to the fact that, at extremely high concentrations, the probability of finding two B atoms in neighbouring lattice sites is not negligible anymore, leading to the formation of partially inactive substitutional B dimers and trimers, long before observing the formation of interstitial dopants or larger precipitates.
First principles simulations provide reliable predictions of the formation energy of different complexes, their electrical activation, and the induced lattice strain. Assuming a random distribution of dopants, they account both for the evolution of hole density and lattice parameter as a function of the dopants concentration, up to the saturation regime, with an extremely limited (or null) diffusion of B during the re-crystallization process. This is favorable to obtain the highest value of B monomer populations, maximizing the hole concentration.  
Finally, we question the impact of the huge ($\sim 4\,m/s$) recrystallization speeds associated to nanosecond laser doping on the microscopic dopant complexes formation, suggesting that the out-of-equilibrium epitaxy further promotes the formation of the energetically favored B trimers.

\section{Methods}
\subsection{Gas Immersion Laser Doping}
Due to the low coherence of the excimer laser, there are no speckles and interference effects degrading the beam energy density. On the other hand, appropriate laser optics allows to strongly reduce the spatial inhomogeneity of the laser energy profile to $1\%$ over a $2\times2\;mm^2$ focused area. The melted layer remains liquid during $20$$-100 ns$ as observed by time resolved reflectivity measurements \cite{chiodi2021}. \\
Each laser shot introduces a fixed surface density, $d_S\approx 1.4\times 10^{14} cm^{-2}$ \cite{desvignes2023,bourguignon1995}.\\
No chlorine atoms brought by the dopant gas, BCl\textsubscript{3}, are observed, in agreement with the Cl segregation coefficient being near zero, as all Cl atoms are expelled when the front reaches the surface\cite{hoummada2012}.\\

\subsection{Hall measurements}
The carrier density was probed by Hall measurements in a square van der Pauw configuration \cite{vanderpauw1958}. A square-like geometry with a central probed surface of $300\times300\;\mu m^2$ was adopted. The p-doped sample is electrically isolated from the n-substrate due to the p-n barrier. In addition, Hall measurements were performed at cryogenic temperatures down to 10 K where the highly resistive substrate is fully insulating. Four Ti (20 nm)/ Au (30nm) contacts were deposited by electron-beam evaporation in an ambient pressure of $10^{-6}\;$mbar and were used as supports for micro-bonding. Indeed, direct micro-bonding on the SiB layers can cause cleavages along the crystalline lines of the high stress induced by B atoms \cite{bonnet2019}. The contacts were isolated from each other by reactive ion etching ($d_{etch}\sim500nm > d_{SiB}$) and were confirmed to be ohmic by measuring the current-voltage curves. Hall measurements were carried-out in both a physical properties measurement system (PPMS) and a home-made set-up. The incident current was fixed at $I = 200\;\mu A$ and perpendicular magnetic field $B_{\perp}$ was swept from $-2$ to $2\;T$. The measured voltage was systematically linear and the hole concentration $h$ could be extracted from \cite{sze2007}:
\begin{equation}\label{eq:Hall}
    h = \gamma\frac{B_{\perp}}{qd(V_H/I)}
\end{equation}
Where q is the electron charge, $d$ the thickness of the layer and $\gamma$ is the Hall mobility factor which is the ratio between the Hall mobility $\mu_H$ and the conductivity mobility $\mu_c$: $\gamma = \frac{\mu_H}{\mu_c}$ . This factor is related to the non parabolicity and the warping of the valence band.
Taking into account the three higher valence bands, Lin et al. \cite{lin1981} have calculated that $\gamma$ decreases slightly with the  hole concentration, and is equal to 0.75 above $10^{18}cm^{-3}$. $\gamma$ has been measured from the mobility for doping levels between $10^{19}cm^{-3}$ and $10^{20}cm^{-3}$ [19]. Thurber et al. [20] obtain $\gamma$=0.7 while Irvin mobility measurements \cite{irvin1962} give $\gamma$=0.8. 
We adopt $\gamma \simeq 0.7$ for our doping range $>10^{19}cm^{-3}$.\\

\subsection{Secondary Ion Mass Spectrometry}
SIMS profiles were done on a 4F Cameca system equipped with a magnetic mass spectrometer. For doping levels above $5\times10^{20}cm^{-3}$, the doping value can be suspected to be modified by matrix effect. This effect was studied in detail by C. Dubois et al. [16]. They show that with an oxygen primary beam, the variation of the relative ion yield is the same for Boron and Silicon so that the matrix effect is corrected by comparing with the silicon signal and using the Relative Sensitivity Factor (RSF). Moreover we analyse only secondary ions at high energy (> 100 eV) which are less sensitive to the chemical surrounding. Boron concentration is obtained by comparison with a standard from the National Institute of Standards (NIST). Depth calibration is obtained with a mechanical profilometer and we take into account the sputter rate difference between the doped layer and silicon substrate. \\
The thickness used to calculate the average atomic and active concentration was determined from the (mainly constant) SIMS profiles (Fig.\ref{fig:fig1}b) as the depth where $\sim\mathrm{95\%}$ of the dose is reached.

\subsection{Scanning Tunnel Electron Microscopy}
For the STEM analysis, a thin lamella ($<100 nm$) is machined vertically in the laser-doped spot thanks to a focus ions beam using a FEI ThermoFisher SCIOS dual beam SEM (UHR NiCol)/FIB (Siderwinder 550 V - 30 kV) with an in-situ easy-lift micromanipulator. The sample was observed in an aberration-corrected FEI ThermoFisher TEM/STEM TitanThemis 200 operating at 200 keV. The convergence halfangle of the probe was 17.6 mrad and the detection inner and outer half-angles for HAADF-STEM were 69 mrad and 200 mrad, respectively. The lamella was imaged along the ⟨1 1 0⟩ zone axis. The micrograph was resolved by 2048 by 2048 pixels. The dwell time was $8 \mu s$ and the total acquisition time 41 s. GPA is performed in Digital Micrograph software on STEM–HAADF images.\\ \\


\subsection{Modeling the hole concentration}
\subsubsection{Computational details}
First principles simulations of B hyperdoped Si are performed by the super-cell method and plane-wave pseudopotential techniques.  After structural optimization we computed the electronic properties of the system by using a 55-Ry cutoff radius for the electronic valence wave function and 440-Ry cutoff radius for the charge density.\\
Super-cells of different sizes are used to compute the formation energy of B complexes as a function of the B concentration as described in Refs [36,66] to simulate Se hyperdoped Si.

\subsubsection{Activation energy}
For a generic defect $D$, the formation energy per B atom as a function of B concentration $x$, in hyperdoped silicon Si$_{1-x}$B$_x$, reads: 
\begin{equation}
\Delta E^{Form.}_{D}(x)=[E_D(x)-N_{Si}\mu_{Si} -N_B\mu_B]/N_B 
\end{equation}
where $E_D(x)$ is the total energy of the supercell with the generic complex $D$, N$_{Si}$ and N$_B$ are the numbers of Si and B atoms in the super-cell, respectively, while the chemical potentials  $\mu_{Si}$  and $\mu_B$  correspond to the ones of Si in bulk silicon and of B in bulk SiB$_6$
at equilibrium with bulk Si.
Our choice of $\mu_B$ corresponds to the chemical potential of B when a phase separation between S and SiB$_6$ occurs. However, since the choice of $\mu_B$ corresponds  to the (arbitrary) choice of zero in the energy scale of Fig.\ref{dft}, this does not affect any of the results presented in the text.

\subsubsection{Hole concentration}
Because of the random nature of the B atoms distribution during the melting process and the high concentrations that are involved ($C_{sat}\sim 9-10\;at.\%$), it can fairly be assumed that there is a non-negligible probability that two or more B atoms locate in neighbouring lattice sites during the crystallisation process. The probability of finding $k\mathrm{-B}$ atoms around a B atom already located on a substitutional site ($\mathrm{B_{Si}}$) is given by the binomial distribution function: 
\begin{equation}\label{eq_binomiale_law}
     p^0_{m} = \left( \begin{array}{c} n \\ k \end{array} \right)\;p^k\;(1-p)^{n-k}
\end{equation}
where $m = k+1$ refers to the total number of B atoms involved in the final B-complex, $n$ is the number of lattice sites, $\left( \begin{array}{c} n \\ k \end{array} \right) = \frac{n!}{k!(n-k)!}$ calculate the various permutations, $p= C_B/ n_{SiB}$ and $(1-p)$ are respectively the probability of picking one B or one Si atom among all the available atoms. For simplicity, $n_{SiB}$ is fixed as $n_{Si}=5\times10^{22}cm^{-3}$, the intrinsic atomic density of c-Si. The probability equations for singlets ($\mathrm{B_{1}}$) and doublets ($\mathrm{B_{2}}$)  are :
\begin{equation}\label{eq_singlet}
    p^0_{1} = \left ( 1 - p\right)^4
\end{equation}
\begin{equation}\label{eq_doublet}
    p^0_{2} = 4p\times\left ( 1 - p\right)^6
\end{equation}
\\
In the case of $p^0_{3}$, the adjustement that reproduces better the trimer concentration simulated assuming a random distribution of substitutional B in the Si lattice is: 
\begin{equation}\label{eq_triplet}
   p^0_{3} = 1.33 \times p_2^0 \times\left[4\left( 1 - p\right)^8+8\left( 1 - p\right)^7 \right]
\end{equation}
In order to both consider the probabilistic formation of $\mathrm{B_m}$ complexes and and their formation energy, we introduce the Boltzmann distribution function with Eq.\ref{eq_binomiale_law}:
\begin{equation}\label{eq_B_distribution}
    p_{m} = \left\{ \begin{array}{ccc} p^0_m\cdot\displaystyle\frac{e^{-\displaystyle\Delta E^{Form}_{D_i}/k_BT} }{Z_m}& & i \leq m \\ 0 & & i > m \\
\end{array}\right.
\end{equation}
where $k_B$ is the Boltzmann constant, $T = 1683K$ the melting temperature of Si and $Z_m = \displaystyle\sum_{D_i}^{i \leq m}e^{-\displaystyle\Delta E^{Form}_{D_i}/k_BT}$ is the partition function that normalizes Eq.\ref{eq_B_distribution} by all the possible B-complexes that involve at most $m$-atoms. For example, considering the complexes that involve at most 2 B atoms, one finds that the $\mathrm{B_{Si}-B_{Si}}$ complex represents on average $\sim 42.5\;\%$ of the doublets, $\mathrm{B2I}\sim 42.5\;\%$ and $\mathrm{B_{Si}-I_B}\sim 15\;\%$. On the other hand, when 3 B atoms are involved, only $0.05\;\%$ takes the form of $\mathrm{B_{Si}-B_{Si}}$ complex while $\sim 90\;\%$ form $\mathrm{B3}$ complexes. The high formation energy of $\mathrm{I_B}$ defects makes them negligible. We write the concentration of each B complex as: 
\begin{equation}\label{eq_Ckp1_proba}
    C_{B_{m}} = C_B\times p_{m}
\end{equation}
Then, knowing the electrical activity of each of the complexes as well as their proportion, it is possible to determine the expected carrier density : 
\begin{equation}\label{eq_active}
    h \simeq C_{B_{1}}+\frac{1}{2}C_{B_{2}}+\frac{1}{3}C_{B_{3}}
\end{equation}
where the fraction $\frac{1}{2}$ holds for the $50\;\%$ of substitutional dimers and $\frac{1}{3}$ for the partial electrical activity of the $\mathrm{B3}$ complex that prevails the other complexes. In this way, we can determine the charge carrier density by determining $C_{B_m}$ from Eq.(\ref{eq_singlet})-(\ref{eq_triplet}).\\
The effect of dopant diffusion can be modelled by considering a random distribution of dopants arranged in the $n$ lattice sites in the neighbour shell to the first B atom.  We consider the nearest neighbour shell S1 (n=4), the next nearest neighbour shell S2 (12 sites for diamond lattice, n=16), and the intermediate case S1.5, with half of the sites of the next nearest neighbor shell (n=10).

\subsection*{Acknowledgements}
FC, GH, LD are grateful for support from the French CNRS RENATECH network, the Physical Measurements Platform of University Paris-Saclay, the French National Research Agency (ANR) under Contract No. ANR-16-CE24-0016-01, ANR-19-CE47-0010-03 and ANR-22-QUA2-0002-02.
AD acknowledges the CINECA award under the ISCRA initiative, for the availability of high-performance computing resources and support (MaCHD-Si project), and R. Colnaghi for technical support on computer hardware. 
GP and BS acknowledge support from the CIR program of French Research Minister.

\bibliography{sn-bibliography}

@article{IRDS2022,
  title={International roadmap for devices and systems},
  year={2022},
  publisher={IEEE}
}

@inproceedings{raghavan2015,
  title={Holisitic device exploration for 7nm node},
  author={Raghavan, Praveen and Bardon, M Garcia and Jang, Doyoung and Schuddinck, Pieter and Yakimets, Dmitry and Ryckaert, Julien and Mercha, Abdelkarim and Horiguchi, Naoto and Collaert, Nadine and Mocuta, Anda and others},
  booktitle={2015 IEEE Custom Integrated Circuits Conference (CICC)},
  pages={1--5},
  year={2015},
  organization={IEEE}
}

@article{Jiang2022,
author = {Jiang, Hao and Wang, Mao and Fu, Jintao and Li, Zhancheng and Shaikh, Mohd Saif and Li, YunJie and Nie, Changbin and Sun, Feiying and Tang, Linlong and Yang, Jun and Qin, Tianshi and Zhou, Dahua and Shen, Jun and Sun, Jiuxun and Feng, Shuanglong and Zhu, Meng and Kentsch, Ulrich and Zhou, Shengqiang and Shi, Haofei and Wei, Xingzhan},
title = {Ultrahigh Photogain Short-Wave Infrared Detectors Enabled by Integrating Graphene and Hyperdoped Silicon},
journal = {ACS Nano},
volume = {16},
number = {8},
pages = {12777-12785},
year = {2022},
doi = {10.1021/acsnano.2c04704},
    note ={PMID: 35900823},
URL = {  
        https://doi.org/10.1021/acsnano.2c04704
},
eprint = { 
        https://doi.org/10.1021/acsnano.2c04704
}
}

@article{hoummada2023,
  title={Analysis of superconducting silicon epilayers by atom probe tomography: composition and evaporation field},
  author={Khalid Hoummada et al.},
  journal={Eur. Phys. J. Appl. Phys. },
  volume={98},
  number={40},
  year={2023},
  publisher={}
}

@article{Bustarret2006,
  title={Superconductivity in doped cubic silicon},
  author={Bustarret, Etienne and Marcenat, C and Achatz, P and Ka{\v{c}}mar{\v{c}}ik, J and L{\'e}vy, F and Huxley, A and Ort{\'e}ga, L and Bourgeois, E and Blase, Xavier and D{\'e}barre, D and others},
  journal={Nature},
  volume={444},
  number={7118},
  pages={465--468},
  year={2006},
  publisher={Nature Publishing Group},
  doi={10.1109/EDL.1986.26429}
}

@article{Nath2024,
author = {Nath, Shimul Kanti and Turan, Ibrahim and Desvignes, Léonard and Largeau, Ludovic and Mauguin, Olivia and Túnica, Marc and Amato, Michele and Renard, Charles and Hallais, Géraldine and Débarre, Dominique and Chiodi, Francesca},
title = {Tuning Superconductivity in Nanosecond Laser-Annealed Boron-Doped Si1–xGex Epilayers},
journal = {physica status solidi (a)},
volume = {221},
number = {24},
pages = {2400313},
doi = {https://doi.org/10.1002/pssa.202400313},
url = {https://onlinelibrary.wiley.com/doi/abs/10.1002/pssa.202400313},
eprint = {https://onlinelibrary.wiley.com/doi/pdf/10.1002/pssa.202400313},
year = {2024}
}

@article{Poumirol2021,
author = {Poumirol, Jean-Marie and Majorel, Cl{\'e}ment and Chery, Nicolas and Girard, Christian and Wiecha, Peter R. and Mallet, Nicolas and Monflier, Richard and Larrieu, Guilhem and Cristiano, Filadelfo and Royet, Anne-Sophie and Alba, Pablo Acosta and Kerdiles, S{\'e}bastien and Paillard, Vincent and Bonafos, Caroline},
title = {Hyper-Doped Silicon Nanoantennas and Metasurfaces for Tunable Infrared Plasmonics},
journal = {ACS Photonics},
volume = {8},
number = {5},
pages = {1393-1399},
year = {2021},
doi = {10.1021/acsphotonics.1c00019},

URL = { 
    
        https://doi.org/10.1021/acsphotonics.1c00019
    
    

},
eprint = { 
    
        https://doi.org/10.1021/acsphotonics.1c00019
    
    

}

}

@article{chiodi2014,
  title={Laser doping for ohmic contacts in n-type Ge},
  author={Chiodi, F and Chepelianskii, AD and Gard{\`e}s, C and Hallais, G and Bouchier, D and D{\'e}barre, D},
  journal={Applied Physics Letters},
  volume={105},
  number={24},
  year={2014},
  publisher={AIP Publishing}
}

@article{chiodi2017,
  title={Proximity-induced superconductivity in all-silicon superconductor/normal-metal junctions},
  author={Chiodi, F and Duvauchelle, J-E and Marcenat, C and D{\'e}barre, D and Lefloch, Fran{\c{c}}ois},
  journal={Physical Review B},
  volume={96},
  number={2},
  pages={024503},
  year={2017},
  publisher={APS}
}

@article{gallacher2012,
  title={Ohmic contacts to n-type germanium with low specific contact resistivity},
  author={Gallacher, Kevin and Velha, Philippe and Paul, Douglas J and MacLaren, I and Myronov, Maksym and Leadley, David R},
  journal={Applied Physics Letters},
  volume={100},
  number={2},
  year={2012},
  publisher={AIP Publishing}
}

@inproceedings{everaert2017,
  title={Sub-10- 9 $\Omega${\textperiodcentered} cm 2 contact resistivity on p-SiGe achieved by Ga doping and nanosecond laser activation},
  author={Everaert, Jean-Luc and Schaekers, Marc and Yu, Hongyu and Wang, L-L and Hikavyy, Andriy and Date, L and del Agua Borniquel, J and Hollar, K and Khaja, FA and Aderhold, W and others},
  booktitle={2017 Symposium on VLSI Technology},
  pages={T214--T215},
  year={2017},
  organization={IEEE}
}

@inproceedings{hung2018,
  title={Novel solutions to enable contact resistivity< 1E-9 $\Omega$-cm 2 for 5nm node and beyond},
  author={Hung, Raymond and Khaja, Fareen Adeni and Hollar, Kelly E and Rao, KV and Munnangi, Samuel and Chen, Yongmei and Okazaki, Motoya and Huang, Yi-Chiau and Li, Xuebin and Chung, Hua and others},
  booktitle={2018 International Symposium on VLSI Technology, Systems and Application (VLSI-TSA)},
  pages={1--2},
  year={2018},
  organization={IEEE}
}

@inproceedings{wang2017,
  title={Comprehensive study of Ga activation in Si, SiGe and Ge with 5$\times$ 10- 10 $\Omega${\textperiodcentered} cm 2 contact resistivity achieved on Ga doped Ge using nanosecond laser activation},
  author={Wang, Lin-Lin and Yu, Hao and Schaekers, Marc and Everaert, J-L and Franquet, Alexis and Douhard, Bastien and Date, L and del Agua Borniquel, J and Hollar, K and Khaja, FA and others},
  booktitle={2017 IEEE International Electron Devices Meeting (IEDM)},
  pages={22--4},
  year={2017},
  organization={IEEE}
}

@article{young1978a,
  title={Laser annealing of boron-implanted silicon},
  author={Young, RT and White, CW and Clark, GJ and Narayan, J and Christie, WH and Murakami, M and King, PW and Kramer, SD},
  journal={Applied Physics Letters},
  volume={32},
  number={3},
  pages={139--141},
  year={1978},
  publisher={American Institute of Physics}
}

@article{kerrien2002,
  title={Ultra-shallow, super-doped and box-like junctions realized by laser-induced doping},
  author={Kerrien, G and Boulmer, J and D{\'e}barre, D and Bouchier, D and Grouillet, A and Lenoble, D},
  journal={Applied surface science},
  volume={186},
  number={1-4},
  pages={45--51},
  year={2002},
  publisher={Elsevier}
}

@article{milazzo2020,
  title={P-type doping of Ge by Al ion implantation and pulsed laser melting},
  author={Milazzo, R and Linser, M and Impellizzeri, G and Scarpa, D and Giarola, M and Sanson, A and Mariotto, G and Andrighetto, A and Carnera, A and Napolitani, E},
  journal={Applied Surface Science},
  volume={509},
  pages={145230},
  year={2020},
  publisher={Elsevier}
}

@article{young1978b,
  title={Laser annealing of diffusion-induced imperfections in silicon},
  author={Young, RT and Narayan, J},
  journal={Applied Physics Letters},
  volume={33},
  number={1},
  pages={14--16},
  year={1978},
  publisher={American Institute of Physics}
}

@article{williams1982,
  title={Metastable doping behavior in antimony-implanted (100) silicon},
  author={Williams, JS and Short, KT},
  journal={Journal of Applied Physics},
  volume={53},
  number={12},
  pages={8663--8667},
  year={1982},
  publisher={American Institute of Physics}
}

@article{huang2005,
  title={Germanium n+/ p junction formation by laser thermal process},
  author={Huang, Jidong and Wu, Nan and Zhang, Qingchun and Zhu, Chunxiang and Tay, Andrew AO and Chen, Guoxin and Hong, Minghui},
  journal={Applied Physics Letters},
  volume={87},
  number={17},
  year={2005},
  publisher={AIP Publishing}
}

@article{hartmann2020,
  title={Ultra-high boron doping of Si and Ge for nanoelectronics and photonics},
  author={Hartmann, Jean-Michel and Frauenrath, Marvin and Richy, J{\'e}r{\^o}me and Veillerot, Marc},
  journal={ECS Transactions},
  volume={98},
  number={5},
  pages={203},
  year={2020},
  publisher={IOP Publishing}
}

@article{rebohle2019,
  title={Flash lamp annealing},
  author={Rebohle, Lars and Prucnal, Slawomir and Reichel, Denise},
  journal={From Basics to Applications/by Lars Rebohle, Slawomir Prucnal, Denise Reichel},
  year={2019},
  publisher={Springer}
}

@article{li2017,
  title={Sulfur-doped silicon photodiode by ion implantation and femtosecond laser annealing},
  author={Li, Chun-Hao and Zhao, Ji-Hong and Yu, Xin-Yue and Chen, Qi-Dai and Feng, Jing and Han, Pei-De and Sun, Hong-Bo},
  journal={IEEE Sensors Journal},
  volume={17},
  number={8},
  pages={2367--2371},
  year={2017},
  publisher={IEEE}
}

@article{winkler2011,
  title={Insulator-to-metal transition in sulfur-doped silicon},
  author={Winkler, Mark T and Recht, Daniel and Sher, Meng-Ju and Said, Aurore J and Mazur, Eric and Aziz, Michael J},
  journal={Physical review letters},
  volume={106},
  number={17},
  pages={178701},
  year={2011},
  publisher={APS}
}

@article{winkler2012,
  title={Studying femtosecond-laser hyperdoping by controlling surface morphology},
  author={Winkler, Mark T and Sher, Meng-Ju and Lin, Yu-Ting and Smith, Matthew J and Zhang, Haifei and Grade{\v{c}}ak, Silvija and Mazur, Eric},
  journal={Journal of Applied Physics},
  volume={111},
  number={9},
  year={2012},
  publisher={AIP Publishing}
}

@article{gandhi2020a,
  title={Chalcogen-hyperdoped germanium for short-wavelength infrared photodetection},
  author={Gandhi, Hemi H and Pastor, David and Tran, Tuan T and Kalchmair, Stefan and Smillie, Lachlan A and Mailoa, Jonathan P and Milazzo, Ruggero and Napolitani, Enrico and Loncar, Marko and Williams, James S and others},
  journal={AIP Advances},
  volume={10},
  number={7},
  year={2020},
  publisher={AIP Publishing}
}

@article{zhou2015a,
  title={Hyperdoping silicon with selenium: solid vs. liquid phase epitaxy},
  author={Zhou, Shengqiang and Liu, Fang and Prucnal, S and Gao, Kun and Khalid, M and Baehtz, C and Posselt, M and Skorupa, W and Helm, M},
  journal={Scientific reports},
  volume={5},
  number={1},
  pages={8329},
  year={2015},
  publisher={Nature Publishing Group UK London}
}

@article{wang2018,
  title={Extended infrared photoresponse in Te-hyperdoped Si at room temperature},
  author={Wang, Mao and Berenc{\'e}n, Y and Garc{\'\i}a-Hemme, E and Prucnal, S and H{\"u}bner, R and Yuan, Ye and Xu, Chi and Rebohle, L and B{\"o}ttger, R and Heller, R and others},
  journal={Physical Review Applied},
  volume={10},
  number={2},
  pages={024054},
  year={2018},
  publisher={APS}
}

@article{wang2019a,
  title={Thermal stability of Te-hyperdoped Si: Atomic-scale correlation of the structural, electrical, and optical properties},
  author={Wang, Mao and H{\"u}bner, R and Xu, Chi and Xie, Yufang and Berenc{\'e}n, Y and Heller, R and Rebohle, L and Helm, M and Prucnal, S and Zhou, Shengqiang},
  journal={Physical Review Materials},
  volume={3},
  number={4},
  pages={044606},
  year={2019},
  publisher={APS}
}

@article{wang2019b,
  title={Breaking the doping limit in silicon by deep impurities},
  author={Wang, Mao and Debernardi, A and Berenc{\'e}n, Y and Heller, R and Xu, Chi and Yuan, Ye and Xie, Yufang and B{\"o}ttger, R and Rebohle, L and Skorupa, W and others},
  journal={Physical Review Applied},
  volume={11},
  number={5},
  pages={054039},
  year={2019},
  publisher={APS}
}

@article{wang2020,
  title={Critical behavior of the insulator-to-metal transition in Te-hyperdoped Si},
  author={Wang, Mao and Debernardi, A and Zhang, Wenxu and Xu, Chi and Yuan, Ye and Xie, Yufang and Berenc{\'e}n, Y and Prucnal, S and Helm, M and Zhou, Shengqiang},
  journal={Physical Review B},
  volume={102},
  number={8},
  pages={085204},
  year={2020},
  publisher={APS}
}

@article{yang2019,
  title={Evidence for vacancy trapping in Au-hyperdoped Si following pulsed laser melting},
  author={Yang, Wenjie and Ferdous, N and Simpson, Peter J and Gaudet, JM and Hudspeth, Quentin and Chow, Philippe K and Warrender, Jeffrey M and Akey, Austin J and Aziz, Michael J and Ertekin, E and others},
  journal={APL Materials},
  volume={7},
  number={10},
  year={2019},
  publisher={AIP Publishing}
}

@article{gandhi2020b,
  title={Gold-hyperdoped germanium with room-temperature sub-band-gap optoelectronic response},
  author={Gandhi, Hemi H and Pastor, David and Tran, Tuan T and Kalchmair, Stefan and Smilie, LA and Mailoa, Jonathan P and Milazzo, Ruggero and Napolitani, Enrico and Loncar, Marco and Williams, James S and others},
  journal={Physical Review Applied},
  volume={14},
  number={6},
  pages={064051},
  year={2020},
  publisher={APS}
}

@article{dissanayake2022,
  title={Carrier lifetimes in gold--hyperdoped silicon—Influence of dopant incorporation methods and concentration profiles},
  author={Dissanayake, Sashini Senali and Pallat, Nicole O and Chow, Philippe K and Lim, Shao Qi and Liu, Yining and Yue, Qianao and Fiutak, Rhoen and Mathews, Jay and Williams, Jim S and Warrender, Jeffrey M and others},
  journal={APL Materials},
  volume={10},
  number={11},
  year={2022},
  publisher={AIP Publishing}
}

@article{lim2021,
  title={A critical evaluation of Ag-and Ti-hyperdoped Si for Si-based infrared light detection},
  author={Lim, Shao Qi and Akey, AJ and Napolitani, E and Chow, PK and Warrender, JM and Williams, JS},
  journal={Journal of Applied Physics},
  volume={129},
  number={6},
  year={2021},
  publisher={AIP Publishing}
}

@article{dong2023,
  title={Nanosecond-laser hyperdoping of intrinsic silicon to modify its electrical and optical properties},
  author={Dong, GY and Yang, HW and Zeng, SJ and Shi, ZQ and Ma, YJ and Wen, C and Yang, WB},
  journal={Optics \& Laser Technology},
  volume={164},
  pages={109517},
  year={2023},
  publisher={Elsevier}
}

@article{wen2021,
  title={Zinc-hyperdoped silicon nanocrystalline layers prepared via nanosecond laser melting for broad light absorption},
  author={Wen, C and Shi, ZQ and Wang, ZJ and Wang, JX and Yang, YJ and Ma, YJ and Yang, WB},
  journal={Optics \& Laser Technology},
  volume={144},
  pages={107415},
  year={2021},
  publisher={Elsevier}
}

@article{pandey1988,
  title={Annealing of heavily arsenic-doped silicon: electrical deactivation and a new defect complex},
  author={Pandey, KC and Erbil, A and Cargill III, GS and Boehme, RF and Vanderbilt, David},
  journal={Physical review letters},
  volume={61},
  number={11},
  pages={1282},
  year={1988},
  publisher={APS}
}

@article{mueller2003,
  title={Arsenic deactivation in Si: Electronic structure and charge states of vacancy-impurity clusters},
  author={Mueller, D Christoph and Alonso, Eduardo and Fichtner, Wolfgang},
  journal={Physical Review B},
  volume={68},
  number={4},
  pages={045208},
  year={2003},
  publisher={APS}
}

@article{mueller2004,
  title={Highly n-doped silicon: Deactivating defects of donors},
  author={Mueller, D Christoph and Fichtner, Wolfgang},
  journal={Physical Review B},
  volume={70},
  number={24},
  pages={245207},
  year={2004},
  publisher={APS}
}

@article{voyles2003,
  title={Imaging individual atoms inside crystals with ADF-STEM},
  author={Voyles, PM and Grazul, JL and Muller, DA},
  journal={Ultramicroscopy},
  volume={96},
  number={3-4},
  pages={251--273},
  year={2003},
  publisher={Elsevier}
}

@article{voyles2002,
  title={Atomic-scale imaging of individual dopant atoms and clusters in highly n-type bulk Si},
  author={Voyles, PM and Muller, DA and Grazul, JL and Citrin, PH and Gossmann, H-JL},
  journal={Nature},
  volume={416},
  number={6883},
  pages={826--829},
  year={2002},
  publisher={Nature Publishing Group UK London}
}

@article{chadi1997,
  title={Fermi-level-pinning defects in highly n-doped silicon},
  author={Chadi, DJ and Citrin, PH and Park, CH and Adler, DL and Marcus, MA and Gossmann, H-J},
  journal={Physical review letters},
  volume={79},
  number={24},
  pages={4834},
  year={1997},
  publisher={APS}
}

@article{debernardi2021,
  title={First principles simulations of microscopic mechanisms responsible for the drastic reduction of electrical deactivation defects in Se hyperdoped silicon},
  author={Debernardi, Alberto},
  journal={Physical Chemistry Chemical Physics},
  volume={23},
  number={43},
  pages={24699--24710},
  year={2021},
  publisher={Royal Society of Chemistry}
}

@article{debernardi2022,
  title={Engineering the insulator-to-metal transition by tuning the population of dopant defects: first principles simulations of Se hyperdoped Si},
  author={Debernardi, Alberto},
  journal={Semiconductor Science and Technology},
  volume={38},
  number={1},
  pages={014002},
  year={2022},
  publisher={IOP Publishing}
}

@phdthesis{desvignes2023,
  title={Laser doped superconducting silicon: from the material to the devices},
  author={Desvignes, L{\'e}onard},
  year={2023},
  school={Universit{\'e} Paris-Saclay}
}

@article{vailionis1999,
  title={Electrically active and inactive B lattice sites in ultrahighly B doped Si (001): An X-ray near-edge absorption fine-structure and high-resolution diffraction study},
  author={Vailionis, A and Glass, G and Desjardins, P and Cahill, David G and Greene, JE},
  journal={Physical review letters},
  volume={82},
  number={22},
  pages={4464},
  year={1999},
  publisher={APS}
}

@article{glass2000,
  title={Ultrahigh B doping (<\~{} 10 22 cm- 3) during Si (001) gas-source molecular-beam epitaxy: B incorporation, electrical activation, and hole transport},
  author={Glass, G and Kim, H and Desjardins, Patrick and Taylor, N and Spila, T and Lu, Q and Greene, JE},
  journal={Physical Review B},
  volume={61},
  number={11},
  pages={7628},
  year={2000},
  publisher={APS}
}

@article{luo2003,
  title={Understanding ultrahigh doping: The case of boron in silicon},
  author={Luo, Xuan and Zhang, SB and Wei, Su-Huai},
  journal={Physical review letters},
  volume={90},
  number={2},
  pages={026103},
  year={2003},
  publisher={APS}
}

@incollection{chiodi2021,
  title={Laser ultra-doped silicon: Superconductivity and applications},
  author={Chiodi, Francesca and Daubriac, Richard and Kerdil{\`e}s, S{\'e}bastien},
  booktitle={Laser Annealing Processes in Semiconductor Technology},
  pages={357--400},
  year={2021},
  publisher={Elsevier}
}

@article{hallais2023,
  title={STEM analysis of deformation and B distribution In nanosecond laser ultra-doped Si1- x B x},
  author={Hallais, G{\'e}raldine and Patriarche, Gilles and Desvignes, L{\'e}onard and D{\'e}barre, Dominique and Chiodi, Francesca},
  journal={Semiconductor Science and Technology},
  volume={38},
  number={3},
  pages={034003},
  year={2023},
  publisher={IOP Publishing}
}

@article{kodera1963,
  title={Diffusion coefficients of impurities in silicon melt},
  author={Kodera, Hiroshi},
  journal={Japanese journal of applied physics},
  volume={2},
  number={4},
  pages={212},
  year={1963},
  publisher={IOP Publishing}
}

@article{wood1981,
  title={Macroscopic theory of pulsed-laser annealing. II. Dopant diffusion and segregation},
  author={Wood, RF and Kirkpatrick, JR and Giles, GE},
  journal={Physical Review B},
  volume={23},
  number={10},
  pages={5555},
  year={1981},
  publisher={APS}
}

@article{bourguignon1995,
  title={Laser modifications of Si (100): Cl surfaces induced by surface melting: etching and cleaning},
  author={Bourguignon, B and Stoica, M and Dragnea, B and Carrez, S and Boulmer, J and Budin, J-P and D{\'e}barre, D and Aliouchouche, A},
  journal={Surface science},
  volume={338},
  number={1-3},
  pages={94--110},
  year={1995},
  publisher={Elsevier}
}

@article{hoummada2012,
  title={Absence of boron aggregates in superconducting silicon confirmed by atom probe tomography},
  author={Hoummada, Khalid and Dahlem, Franck and Kociniewski, Thierry and Boulmer, Jacques and Dubois, Christiane and Prudon, Gilles and Bustarret, Etienne and Courtois, Herv{\'e} and D{\'e}barre, Dominique and Mangelinck, Dominique},
  journal={Applied Physics Letters},
  volume={101},
  number={18},
  year={2012},
  publisher={AIP Publishing}
}

@article{vanderpauw1958,
  title={A method of measuring the resistivity and Hall coefficient on lamellae of arbitrary shape},
  author={van der Pauw, Leo J},
  journal={Philips technical review},
  volume={20},
  pages={220--224},
  year={1958}
}

@phdthesis{bonnet2019,
  title={Mesures r{\'e}sonantes des propri{\'e}t{\'e}s hautes fr{\'e}quences du silicium supraconducteur ultra-dop{\'e} au bore par laser},
  author={Bonnet, Pierre},
  year={2019},
  school={Universit{\'e} Paris Saclay (COmUE)}
}

@book{sze2007,
  title={Physics of semiconductor devices, third edition},
  author={Sze, Simon M and and Ng, Kwok K},
  year={2007},
  publisher={John wiley \& sons}
}

@article{irvin1962,
  title={Resistivity of bulk silicon and of diffused layers in silicon},
  author={Irvin, John C},
  journal={Bell System Technical Journal},
  volume={41},
  number={2},
  pages={387--410},
  year={1962},
  publisher={Wiley Online Library}
}

@article{lin1981,
  title={Theoretical analysis of hall factor and hall mobility in p-type silicon},
  author={Lin, JF and Li, SS and Linares, LC and Teng, KW},
  journal={Solid-State Electronics},
  volume={24},
  number={9},
  pages={827--833},
  year={1981},
  publisher={Elsevier}
}

@article{solmi1990,
  title={High-concentration boron diffusion in silicon: Simulation of the precipitation phenomena},
  author={Solmi, S and Landi, E and Baruffaldi, F},
  journal={Journal of applied physics},
  volume={68},
  number={7},
  pages={3250--3258},
  year={1990},
  publisher={American Institute of Physics}
}

@inproceedings{earles2004,
  title={Formation of ultrashallow junctions in 500 eV boron implanted silicon using nonmelt laser annealing},
  author={Earles, Susan and Law, ME and Jones, KS and Frazer, J and Talwar, Somit and Downey, Dan and Arevalo, Edwin},
  booktitle={12th IEEE International Conference on Advanced Thermal Processing of Semiconductors, 2004. RTP 2004.},
  pages={143--147},
  year={2004},
  organization={IEEE}
}

@article{cristiano2016,
  title={Defect evolution and dopant activation in laser annealed Si and Ge},
  author={Cristiano, Fuccio and Shayesteh, M and Duffy, R and Huet, K and Mazzamuto, F and Qiu, Yang and Quillec, M and Henrichsen, HH and Nielsen, PF and Petersen, Dirch Hjorth and others},
  journal={Materials Science in Semiconductor Processing},
  volume={42},
  pages={188--195},
  year={2016},
  publisher={Elsevier}
}

@article{sharp2006,
  title={Deactivation of ultrashallow boron implants in preamorphized silicon after nonmelt laser annealing with multiple scans},
  author={Sharp, JA and Cowern, NEB and Webb, RP and Kirkby, KJ and Giubertoni, Damiano and Gennaro, Salvatore and Bersani, Massimo and Foad, MA and Cristiano, F and Fazzini, PF},
  journal={Applied physics letters},
  volume={89},
  number={19},
  year={2006},
  publisher={AIP Publishing}
}

@article{jain2004,
  title={Metastable boron active concentrations in Si using flash assisted solid phase epitaxy},
  author={Jain, SH and Griffin, PB and Plummer, JD and Mccoy, S and Gelpey, J and Selinger, T and Downey, DF},
  journal={Journal of applied physics},
  volume={96},
  number={12},
  pages={7357--7360},
  year={2004},
  publisher={AIP Publishing}
}

@article{yeong2008,
  title={Understanding of boron junction stability in preamorphized silicon after optimized flash annealing},
  author={Yeong, SH and Colombeau, B and Poon, CH and Mok, KRC and See, A and Benistant, F and Tan, DXM and Pey, KL and Ng, CM and Chan, L and others},
  journal={Journal of the Electrochemical Society},
  volume={155},
  number={7},
  pages={H508},
  year={2008},
  publisher={IOP Publishing}
}

@article{nishikawa2011,
  title={Electrical properties of polycrystalline silicon films formed from amorphous silicon films by flash lamp annealing},
  author={Nishikawa, Takuya and Ohdaira, Keisuke and Matsumura, Hideki},
  journal={Current Applied Physics},
  volume={11},
  number={3},
  pages={604--607},
  year={2011},
  publisher={Elsevier}
}

@article{do2014,
  title={Effect of flash lamp annealing on electrical activation in boron-implanted polycrystalline Si thin films},
  author={Do, Woori and Jin, Won-Beom and Choi, Jungwan and Bae, Seung-Muk and Kim, Hyoung-June and Kim, Byung-Kuk and Park, Seungho and Hwang, Jin-Ha},
  journal={Materials Research Bulletin},
  volume={58},
  pages={164--168},
  year={2014},
  publisher={Elsevier}
}

@article{jain2002,
  title={Transient enhanced diffusion of boron in Si},
  author={Jain, SC and Schoenmaker, Wim and Lindsay, Richard and Stolk, PA and Decoutere, Stefaan and Willander, Magnus and Maes, HE},
  journal={Journal of applied physics},
  volume={91},
  number={11},
  pages={8919--8941},
  year={2002},
  publisher={American Institute of Physics}
}

@incollection{marques2021,
  title={Atomistic modeling of laser-related phenomena},
  author={Marqu{\'e}s, Luis A and Aboy, Mar{\'\i}a and L{\'o}pez, Pedro and Santos, Iv{\'a}n and Pelaz, Lourdes and Fisicaro, Giuseppe},
  booktitle={Laser Annealing Processes in Semiconductor Technology},
  pages={79--136},
  year={2021},
  publisher={Elsevier}
}

@article{aboy2014,
  title={Modeling of defects, dopant diffusion and clustering in silicon},
  author={Aboy, Maria and Santos, Iv{\'a}n and Pelaz, Lourdes and Marqu{\'e}s, Luis A and L{\'o}pez, Pedro},
  journal={Journal of Computational Electronics},
  volume={13},
  pages={40--58},
  year={2014},
  publisher={Springer}
}

@article{zhou2015b,
  title={Comparative study on the localized surface plasmon resonance of boron-and phosphorus-doped silicon nanocrystals},
  author={Zhou, Shu and Pi, Xiaodong and Ni, Zhenyi and Ding, Yi and Jiang, Yingying and Jin, Chuanhong and Delerue, Christophe and Yang, Deren and Nozaki, Tomohiro},
  journal={Acs Nano},
  volume={9},
  number={1},
  pages={378--386},
  year={2015},
  publisher={ACS Publications}
}

@article{patra2019,
  title={Electrically active boron doping in the core of Si nanocrystals by planar inductively coupled plasma CVD},
  author={Patra, Chandralina and Das, Debajyoti},
  journal={Journal of Applied Physics},
  volume={126},
  number={15},
  year={2019},
  publisher={AIP Publishing}
}

@article{hunter2019,
  title={Probing dopant locations in silicon nanocrystals via high energy X-ray diffraction and reverse monte carlo simulation},
  author={Hunter, Katharine I and Bedford, Nicholas and Schramke, Katelyn and Kortshagen, Uwe R},
  journal={Nano letters},
  volume={20},
  number={2},
  pages={852--859},
  year={2019},
  publisher={ACS Publications}
}

@article{oliva2016,
  title={Doping silicon nanocrystals and quantum dots},
  author={Oliva-Chatelain, Brittany L and Ticich, Thomas M and Barron, Andrew R},
  journal={Nanoscale},
  volume={8},
  number={4},
  pages={1733--1745},
  year={2016},
  publisher={Royal Society of Chemistry}
}

@article{bisognin2007,
  title={Substitutional B in Si: Accurate lattice parameter determination},
  author={Bisognin, Gabriele and De Salvador, Davide and Napolitani, Enrico and Berti, Marina and Carnera, Alberto and Mirabella, S and Romano, L and Grimaldi, MG and Priolo, F},
  journal={Journal of applied physics},
  volume={101},
  number={9},
  year={2007},
  publisher={AIP Publishing}
}

@article{bisognin2006a,
  title={Lattice strain induced by boron clusters in crystalline silicon},
  author={Bisognin, G and De Salvador, D and Napolitani, E and Carnera, A and Bruno, E and Mirabella, S and Priolo, F and Mattoni, A},
  journal={Semiconductor science and technology},
  volume={21},
  number={6},
  pages={L41},
  year={2006},
  publisher={IOP Publishing}
}

@article{bisognin2006b,
  title={Lattice strain of B--B pairs formed by He irradiation in crystalline Si1- xBx/Si},
  author={Bisognin, G and De Salvador, D and Napolitani, E and Carnera, A and Romano, L and Piro, AM and Mirabella, S and Grimaldi, MG},
  journal={Nuclear Instruments and Methods in Physics Research Section B: Beam Interactions with Materials and Atoms},
  volume={253},
  number={1-2},
  pages={55--58},
  year={2006},
  publisher={Elsevier}
}

@article{sardela1994,
  title={Relation between electrical activation and the B-induced strain in Si determined by reciprocal lattice mapping},
  author={Sardela Jr, MR and Radamson, HH and Ekberg, JO and Sundgren, J-E and Hansson, GV},
  journal={Semiconductor science and technology},
  volume={9},
  number={6},
  pages={1272},
  year={1994},
  publisher={IOP Publishing}
}

@article{baribeau1991,
  title={Characterization of boron-doped silicon epitaxial layers by x-ray diffraction},
  author={Baribeau, J-M and Rolfe, SJ},
  journal={Applied physics letters},
  volume={58},
  number={19},
  pages={2129--2131},
  year={1991},
  publisher={American Institute of Physics}
}

@article{boureau2018,
  title={Lattice contraction due to boron doping in silicon},
  author={Boureau, Victor and Hartmann, Jean Michel and Claverie, Alain},
  journal={Materials Science in Semiconductor Processing},
  volume={87},
  pages={65--68},
  year={2018},
  publisher={Elsevier}
}

@article{dunham2006,
  title={Calculations of effect of anisotropic stress/strain on dopant diffusion in silicon under equilibrium and nonequilibrium conditions},
  author={Dunham, Scott T and Diebel, Milan and Ahn, Chihak and Shih, Chen Luen},
  journal={Journal of Vacuum Science \& Technology B: Microelectronics and Nanometer Structures Processing, Measurement, and Phenomena},
  volume={24},
  number={1},
  pages={456--461},
  year={2006},
  publisher={American Vacuum Society}
}

\end{document}